\documentclass[aps,pre,reprint,twocolumn,showpacs,preprintnumbers,groupedaddress,floatfix,amsmath,amssymb]{revtex4-1}

\usepackage{graphicx}
\usepackage{color}

\newcommand{\physrep}{Phys.~Rep.} % Physics Reports
\newcommand{\apjl}{Astrophys.~J.~Lett.} % The Astrophysical Journal Letters
\newcommand{\araa}{Annu.~Rev.~Astron.~Astrophys.} % Annual Review of Astronomy and Astrophysics
\newcommand{\mnras}{Mon.~Not.~R.~Astron.~Soc.} % Monthly Notices of the Royal Astronomical Society
\newcommand{\Rey}{\text{Re}}
\newcommand{\Rm}{\text{Rm}}

\begin{document}

\title{The small-scale dynamo: Breaking universality at high Mach numbers}

\author{Dominik R.G. Schleicher}
  \email{dschleic@astro.physik.uni-goettingen.de}
  \affiliation{Institut f\"ur Astrophysik, Georg-August-Universit\"at G\"ottingen, Friedrich-Hund-Platz 1, D-37077 G\"ottingen, Germany}
\author{Jennifer Schober}
 \email{schober@stud.uni-heidelberg.de}
 \affiliation{Universit\"at Heidelberg, Zentrum f\"ur Astronomie, Institut f\"ur Theoretische Astrophysik, Albert-\"Uberle-Strasse\ 2, D-69120 Heidelberg, Germany}
 \author{Christoph Federrath}
  \email{christoph.federrath@monash.edu}
  \affiliation{Monash Centre for Astrophysics, School of Mathematical Sciences, Monash University, Vic 3800, Australia \\ Universit\"at Heidelberg, Zentrum f\"ur Astronomie, Institut f\"ur Theoretische Astrophysik, Albert-\"Uberle-Strasse\ 2, D-69120 Heidelberg, Germany}
\author{Stefano Bovino}
  \email{sbovino@astro.physik.uni-goettingen.de}
  \affiliation{Institut f\"ur Astrophysik, Georg-August-Universit\"at G\"ottingen, Friedrich-Hund-Platz 1, D-37077 G\"ottingen, Germany\ }
%\author{Robi Banerjee}
 % \email{banerjee@hs.uni-hamburg.de}
 % \affiliation{UniversitŠt Hamburg, Hamburger Sternwarte, Goejenbergsweg 112, 21029 Hamburg}
\author{Wolfram Schmidt}
  \email{schmidt@astro.physik.uni-goettingen.de}
  \affiliation{Institut f\"ur Astrophysik, Georg-August-Universit\"at G\"ottingen, Friedrich-Hund-Platz 1, D-37077 G\"ottingen, Germany\ \ }

\date{\today}

\begin{abstract}
The small-scale dynamo may play a substantial role in magnetizing the Universe under a large range of conditions, including subsonic turbulence at low Mach numbers, highly supersonic turbulence at high Mach numbers and a large range of magnetic Prandtl numbers Pm, i.e. the ratio of kinetic viscosity to magnetic resistivity. Low Mach numbers may in particular lead to the well-known, incompressible Kolmogorov turbulence, while for high Mach numbers, we are in the highly compressible regime, thus close to Burgers turbulence. In this study, we explore whether in this large range of conditions, a universal behavior can be expected. Our starting point are previous investigations in the kinematic regime. Here, analytic studies based on the Kazantsev model have shown that the behavior of the dynamo depends significantly on Pm and the type of turbulence, and numerical simulations indicate a strong dependence of the growth rate on the Mach number of the flow. Once the magnetic field saturates on the current amplification scale, backreactions occur and the growth is shifted to the next-larger scale. We employ a Fokker-Planck model to calculate the magnetic field amplification during the non-linear regime, and find a resulting power-law growth that depends on the type of turbulence invoked. For Kolmogorov turbulence, we confirm previous results suggesting a linear growth of magnetic energy. For more general turbulent spectra, where the turbulent velocity $v_t$ scales with the characteristic length scale as $u_\ell\propto \ell^{\vartheta}$, we find that the magnetic energy grows as $(t/T_{ed})^{2\vartheta/(1-\vartheta)}$, with $t$ the time-coordinate and $T_{ed}$ the eddy-turnover time on the forcing scale of turbulence. For Burgers turbulence, $\vartheta=1/2$,  a quadratic rather than linear growth may thus be expected, as the spectral energy increases from smaller to larger scales more rapidly. The quadratic growth is due to the initially smaller growth rates obtained for Burgers turbulence, and thus implies longer timescales until saturation is reached. Similarly, we show that the characteristic length scale of the magnetic field grows as $t^{1/(1-\vartheta)}$ in the general case, implying $t^{3/2}$ for Kolmogorov and $t^2$ for Burgers turbulence. Overall, we find that high Mach numbers, as typically associated with steep spectra of turbulence, may break the previously postulated universality, and introduce a dependence on the environment also in the non-linear regime.
\end{abstract}

\maketitle

\section{Introduction}
The small-scale dynamo has been suggested to operate under a large range of different conditions, including the solar surface \citep{Graham09, Graham10}, galaxies and galaxy clusters \citep{Beck96, Kulsrud97, Subramanian06, Shukurov06, Arshakian09, deSouza10}, the intergalactic medium \citep{Ryu08} and the formation of the first stars and galaxies \citep{Schleicher10c, Sur10, Federrath11, Schober12, Sur12, Turk12, Latif12b, Latif12a}. It thus operates on a large range of different conditions, concerning for instance the magnetic Prandtl number Pm, i.e. the ratio of kinematic viscosity $\nu$ to magnetic resistivity $\eta$, the Mach number of the turbulence $\mathcal{M}$, i.e. the ratio of turbulent velocities to the sound speed, and, most likely related to the Mach number, the expected type of turbulence in the system.

Most studies of the small-scale dynamo performed so far have focused on incompressible Kolmogorov turbulence \citep{Kolmogorov41}, assuming a scaling relation $u_\ell\propto \ell^{1/3}$ between turbulent velocity $u_\ell$ and length scale $\ell$. For Kolmogorov turbulence, it was previously concluded that the magnetic energy grows exponentially in the kinematic regime \citep[e.g.][]{Kazantsev68, Subramanian98, Brandenburg05, Schober12b} and linearly once the backreactions from the magnetic field become important (e.g. \citep{Scheko02, Cho09, Beresnyak12}). The latter was interpreted by \citet{Beresnyak12} as evidence for universality of the small-scale dynamo, suggesting that a fixed fraction of the global turbulence dissipation rate is converted into magnetic energy. 

However, observations of turbulence in molecular clouds \citep[e.g.][]{Larson81, Ossenkopf02} and numerical simulations of supersonic turbulence \citep[e.g.][]{Boldyrev02, Kritsuk07, Schmidt08, Federrath10turb} often reveal steeper turbulent spectra, typically inbetween the incompressible Kolmogorov turbulence and the highly compressible Burgers turbulence \citep{Burgers48}. So far, only a small amount of studies have investigated the turbulent dynamo in this regime. For instance, \citet{Haugen04c} provided the first study exploring the dependence of the dynamo on the Mach number in simulations of driven turbulence, and \citet{Balsara01, Balsara04, Balsara05} explored the amplification of magnetic fields in turbulence produced from strong supernova shocks. The first systematic study covering turbulent Mach numbers from $0.02$ to $20$ and different types of turbulence driving has been pursued by \citet{FederrathPRL}, while the effect of a large range of different turbulence spectra has been explored by \citet{Schober12b} based on the Kazantsev model \citep{Kazantsev68}.

{ We note that the small-scale dynamo has also been studied in the context of so-called shell models \citep{Cattaneo09, Lessinnes09a, Lessinnes09b, Plunian12}. The latter originate from shell models of hydrodynamical turbulence, which originally considered turbulence in 2D \citep{Lorenz71, Gledzer73, Desnianskii74}, but were extended to 3D once a description of kinetic helicity was obtained \citep{Kadanoff95}. The first 2D MHD shell model has been derived by \citet{Frik84}, while 3D models have been developed by \citet{Brandenburg96, Basu98, Frick98}. More sophisticated processes such as non-local interactions \citep{Frick06, Plunian07}, anisotropies \citep{Nigro04} and the Hall effect \citep{Frick03} have been incorporated in more recent studies. These approaches allow to study both the evolution of the power spectrum as well as the saturated regime, and are highly complementary to the methods presented here.  }

In the following, we will consider the small-scale dynamo in the kinematic and non-linear regime, and present evidence from existing and new calculations suggesting a strong dependence on the magnetic Prandtl number, as well as the Mach number of the flow. In section 2, we summarize the evidence and indications for a non-universal behavior in the kinematic regime, which has been derived in previous studies. In section 3, we present the first exploration concerning different types of turbulence during the non-linear phase of the dynamo, where the backreaction of the magnetic field becomes important.  We show that a linear growth is only obtained in the case of Kolmogorov turbulence, while steeper power laws result from turbulent spectra with $\vartheta>1/3$. We discuss the physical implications in section 4, and summarize our main results in section 5.

\section{Non-Universality in the kinematic regime}
The small-scale dynamo is well-studied in the kinematic regime, where an exponential growth of the magnetic field is expected on the viscous scale. The growth rate of the magnetic field can be calculated in the framework of the Kazantsev model, assuming homogeneous turbulence that is $\delta$-correlated in time, or with 3-dimensional magneto-hydrodynamical simulations. In this section, we discuss hints and evidence for a non-universal behavior in the kinematic regime.

\subsection{Indications for non-universality in the Kazantsev model}
The amplification of magnetic fields is governed by the induction equation, which is given as\begin{equation}
\partial_t\vec{B}=\nabla\times\vec{v}\times\vec{B}+\eta\Delta\vec{B}.\label{induction}
\end{equation}
{ We assume in the following $\langle \vec{B}\rangle=0$, although we note that scenarios considering  $\langle \vec{B}\rangle\neq0$ have been recently explored by \citet{Boldyrev05, Malyshkin07, Malyshkin09, Malyshkin10}.} In the Kazantsev model, the velocity field and the magnetic field are decomposed into a mean field, denoted with brackets $\langle \rangle$, and a fluctuating component denoted with $\delta$:\begin{equation}
\vec{v}=\langle \vec{v}\rangle+\delta\vec{v},\quad \vec{B}=\langle\vec{B}\rangle +\delta\vec{B}.
\end{equation}
A central input is the correlation function of the turbulent velocity, which is $\delta$-correlated in time and (in the absence of helicity) can be decomposed as\begin{eqnarray}
\langle \delta v_i(\vec{r}_1,t)\delta v_j(\vec{r}_2,s)\rangle &=& T_{ij}(r) \delta(t-s),\\
T_{ij}(r)&=&\left( \delta_{ij}-\frac{r_ir_j}{r^2} \right)T_N(r)+\frac{r_ir_j}{r^2}T_L(r),\nonumber
\end{eqnarray}
with $r=|\vec{r}_1-\vec{r}_2|$ and $T_N$, $T_L$ are the transverse and longitudinal parts of the correlation function, respectively \citep{Batchelor53}. The same definitions can be applied to the magnetic field, yielding a two-point correlation function $M_{ij}(r,t)$ with transverse and longitudinal components $M_N(r,t)$ and $M_L(r,t)$.  Unlike the velocity field, the magnetic field is always divergence-free, leading to the additional constraint
\begin{equation}
  M_{\text{N}} = \frac{1}{2r} \frac{\text{d}}{\text{d} r}\left(r^2M_\text{L}\right).
\label{MN}
\end{equation}
 { As the Kazantsev model assumes that the flow is $\delta$-correlated in time, concepts such as viscosity or the magnetic Prandtl number cannot be directly incorporated into the flow, as the turbulent velocity field is destroyed and regenerated at each instant, leaving no time for viscosity to act. However, it can be indirectly included by adopting turbulent velocity spectra that become steeper below a given viscous scale $\ell_\nu$. This is the approach employed here.} For a given relation of type
\begin{equation}
  u_\ell \propto \ell^{\vartheta}
\label{TurbPower}
\end{equation}
in the inertial range, the longitudinal correlation function of turbulence can be parametrized as~\citep{Schober12b}
\begin{equation}
  T_\text{L}(r) = \begin{cases} 
                     \frac{VL}{3}\left(1-\Rey^{(1-\vartheta)/(1+\vartheta)}\left(\frac{r}{L}\right)^{2}\right)        & 0<r<\ell_\nu \\ 
  		     \frac{VL}{3}\left(1-\left(\frac{r}{L}\right)^{\vartheta+1}\right)                                & \ell_\nu<r<L \\ 
  		     0                                                                                                & L<r, 
  		  \end{cases}
\label{TL}
\end{equation}
with $\ell_\nu$ the viscous scale, $L$ the driving scale of turbulence, $V$ the turbulent velocity on scale $L$, $\mathrm{Re}=VL/\nu$  the Reynolds number of the gas and $\nu$ the kinetic viscosity. Similarly, we have
\begin{equation}
  T_\text{N}(r) = \begin{cases} 
                     \frac{VL}{3}\left(1-\theta(\vartheta)\Rey^{(1-\vartheta)/(1+\vartheta)} \left(\frac{r}{L}\right)^{2}\right)  	& 0<r<\ell_\nu \\
  		     \frac{VL}{3}\left(1-\theta(\vartheta)\left(\frac{r}{L}\right)^{\vartheta+1}\right)                              & \ell_\nu<r<L \\ 
  		     0                                                                                                   	& L<r, 
  	          \end{cases}
\label{TN}
\end{equation}
with $\theta(\vartheta)=(21-38\vartheta)/5$. As we expect an exponential growth of the magnetic energy as a function of time, we make the following ansatz for the kinematic regime:
\begin{equation}
  M_\text{L}(r,t) \equiv \frac{1}{r^2\sqrt{\kappa_\text{diff}}}\psi(r)\text{e}^{2\Gamma t}.\label{sepansatz}
\end{equation}
Inserting (\ref{sepansatz}) in the induction equation (\ref{induction}), one obtains the Kazantsev equation, which is of the same form as the quantum-mechanical Schr\"odinger equation:
\begin{equation}
  -\kappa_\text{diff}(r)\frac{\text{d}^2\psi(r)}{\text{d}^2r} + U(r)\psi(r) = -\Gamma \psi(r).
\label{Kazantsev}
\end{equation}
In this framework, the amplification depends on the effective potential $U(r)$ in Eq.~(\ref{Kazantsev}), which depends on the properties of turbulence via
\begin{eqnarray}
  U(r) &\equiv& \frac{\kappa_\text{diff}''}{2} - \frac{(\kappa_\text{diff}')^2}{4\kappa_\text{diff}} + \frac{2\kappa_\text{diff}}{r^2} + \frac{2T_\text{N}'}{r} + \frac{2(T_\text{L}-T_\text{N})}{r^2},\nonumber\\
  \kappa_{\mathrm{diff}}&=&\eta+T_L(0)-T_L(r).
\label{GeneralPotential}
\end{eqnarray}
{ As recently shown by \citet{Schober12b}, this form of the potential also accounts for the effect of compressibility by keeping terms related to $\nabla\cdot\vec{v}$ during the derivation.} The equation can be solved using the WKB approximation in the limit of $\mathrm{Pm}\rightarrow\infty$ \citep{Kazantsev68, Subramanian98, Brandenburg05, Schober12b}.  For Kolmogorov turbulence, one obtains\begin{equation}
\Gamma_{K,\mathrm{Pm}\,\gg\,1}=1.028\frac{V}{L} \mathrm{Re}^{1/2}.
\end{equation}
In a recent study, analytical solutions based on the WKB approximation have been derived in the limit $\mathrm{Pm}\ll1$ by \citet{Schober12c}. For Kolmogorov, they yield
\begin{equation}
\Gamma_{K,\mathrm{Pm}\,\ll\,1}=0.0268\frac{V}{L} \mathrm{Rm}^{1/2},
\end{equation}
with $\mathrm{Rm}=VL/\eta$ the magnetic Reynolds number, and $\eta$ the magnetic diffusivity. We thus observe a fundamental difference between the limiting cases $\mathrm{Pm}\ll1$ and $\mathrm{Pm}\gg1$ in the kinematic regime:  For $\mathrm{Pm}\gg1$, magnetic field amplification occurs predominantly on the viscous scale, corresponding to the most negative range of the potential. For $\mathrm{Pm}\ll1$, on the other hand, the resistive scale becomes larger than the visous scale. Amplification on the viscous scale is thus not possible, and the strongest contribution is close to the resistive scale due to the short eddy-times. Correspondingly, the growth rate of the magnetic field depends on the Reynolds number Re for $\mathrm{Pm}\gg1$, and on the magnetic Reynolds number Rm for $\mathrm{Pm}\ll1$ \citep[see also][]{Kleeorin12}. 

The results can be generalized further for different types of turbulence. In the limit $\mathrm{Pm}\gg1$, one obtains \citep{Schober12b}\begin{equation}
\Gamma_{\vartheta,\mathrm{Pm}\,\gg\,1}=\frac{(163-304\vartheta)}{60}\frac{V}{L}\mathrm{Re}^{(1-\vartheta)/(1+\vartheta)}.
\end{equation}
In the regime $\mathrm{Pm}\ll1$, one finds a similar relation \citep{Schober12c},
\begin{equation}
  \Gamma = \alpha\frac{V}{L}\Rm^{(1-\vartheta)/(1+\vartheta)}
\label{ansatz}
\end{equation}
with the prefactor $\alpha$ defined through the quantities
\begin{eqnarray}
  a(\vartheta) & = & \vartheta(56-103\vartheta), \\
  b(\vartheta) & = & \vartheta(79-157\vartheta),\\
    c(\vartheta) &=& \frac{25 + {\sqrt{135\,a(\vartheta) + {\left(b(\vartheta) - 25\right)}^2}} - b(\vartheta)}{a(\vartheta)}.
\end{eqnarray}
as
\begin{equation}
  \alpha =\frac{a(\vartheta)}{5}\,c(\vartheta)^{\frac{\vartheta - 1}{1 + \vartheta }}\,\exp\left(\sqrt{\frac{5}{3\,a(\vartheta)}}\,\pi \,\left(\vartheta - 1\right) - 2\right).
\end{equation}
A numerical evaluation shows that these coefficients are smaller by about two orders of magnitude in the limit $\mathrm{Pm}\ll1$, assuming the same type of turbulence. This can be expected, as the amplification then occurs on larger scales, with larger eddy-turnover times.

 Similarly, also the type of turbulence reflected in the parameter $\vartheta$ may change the amplification rate by about an order of magnitude, in case of the same value of Rm. The most efficient amplification rate occurs for Kolmogorov turbulence, $\vartheta=1/3$, while it is less efficient for highly compressible Burgers turbulence, $\vartheta=1/2$, for which the turbulent velocities decrease more rapidly with length scale. 

The Kazantsev model thus indicates that the behavior of the dynamo depends both on the type of turbulence and the magnetic Prandtl number. A potential restriction of the underlying model is the assumption of $\delta$-correlated turbulence, although the characteristic timescales are certainly small compared to the dynamical time. To investigate the resulting uncertainties, we refer the reader to  \citet{Schekochihin01}. The main results from these considerations are thus the following:\begin{itemize}
\item The behavior of the small-scale dynamo depends sensitively on the value of Pm, and in particular whether $\mathrm{Pm}\ll1$ or $\mathrm{Pm}\gg1$. We note that there is a continuous transition at $\mathrm{Pm}\sim1$, as detailed by \citet{Bovino12}.
\item The adopted type of turbulence has a significant influence on the efficiency of magnetic field amplification, as turbulent spectra with $\vartheta>1/3$ correspond to larger eddy-turnover times and smaller amplification rates.
\end{itemize}

\subsection{Results from numerical simulations}

Due to the numerical viscosity and resistivity, it is difficult to perform magneto-hydrodynamical simulations with Pm significantly different from 1. However, a limited range of Pm has nevertheless been explored. For instance, \citet{Haugen04a} investigated magnetic Prandtl numbers between $0.1$ and $30$. For $\mathrm{Pm}<1$, they report that the miminum magnetic Reynolds number required for dynamo action, Rm$_c$, scales as\begin{equation}
\mathrm{Rm}_c\sim35\pi\,\mathrm{Pm}^{-1/2}.
\end{equation}
We note that the factor $\pi$ in the above is due to their definition of the magnetic Reynolds number. They further report differences in the obtained power spectra, indicating a steeper decrease on small scales for small values of Pm. 

\citet{Scheko07} and \citet{Iskakov07} report numerical simulations exploring the small-scale dynamo from Pm$\sim0.017$ up to $\mathrm{Pm}=1$. In this regime, they find that even for constant values of Rm, the growth rate decreases with decreasing Pm. In particular, for $\mathrm{Pm}\sim1$ and $\mathrm{Rm}\sim830$, they report a normalized growth rate of $1.8$, which decreases to $0.9$ for $\mathrm{Pm}\sim0.2$ and the same magnetic Reynolds number. { The simulations further indicate that the value of Rm$_c$ settles to a constant limit for $\mathrm{Pm}\ll1$, $\mathrm{Re}\gg1$ and $\mathrm{Rm}\gg1$, even though this case is hard to numerically explore. The scaling of the growth rate on Rm, on the other hand, has not been conclusively explored.}

All in all, simulations thus show that the growth rate depends on the magnetic Prandtl number even in the range $\mathrm{Pm}\lesssim1$. Another quantity which was shown to influence the dynamo is the Mach number of the gas. \citet{Haugen04c} explored Mach numbers in the range of $0.1-2.1$ and reported a clear dependence of the critical magnetic Reynolds number for dynamo action on the Mach number $\mathcal{M}$. For $\mathrm{Pm}\sim5$, they report $\mathrm{Rm}_c\sim25\pi$ for $\mathcal{M}<1$ and a rapid increase to $\mathrm{Rm}_c\sim45\pi$ for $\mathcal{M}>1$. A similar behavior was found for $\mathrm{Pm}\sim1$, with critical values of $\sim40\pi$ and $\sim80\pi$, respectively.

\begin{figure}[htbp]
\begin{center}
\includegraphics[scale=0.51]{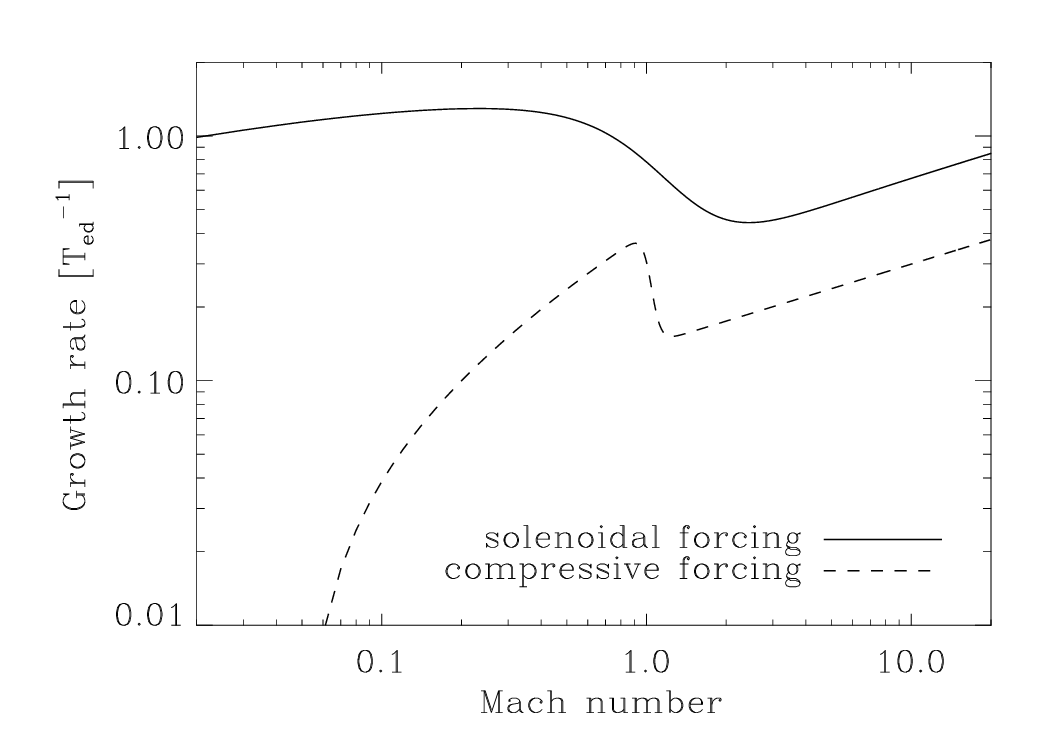}
\caption{A fit to the growth rates obtained by \citet{FederrathPRL} for compressive and solenoidal forcing as a function of Mach number. { These simulations correspond to $\mathrm{Pm}\sim2$.}}
\label{fig_highmach}
\end{center}
\end{figure}

A larger series of simulations has been reported by \citet{FederrathPRL}, exploring Mach numbers from $0.02$ up to $20$, with compressive and solenoidal forcing, respectively. They performed a fit to the growth rate and saturation levels as a function of Mach number, using the function\begin{equation}
f(\mathcal{M})=\left( p_0\frac{\mathcal{M}^{p_1}+p_2}{\mathcal{M}^{p_3}+p_4}+p_5\right)\mathcal{M}^{p_6}.
\end{equation}
The fit coefficients for the different cases are given in Table~\ref{table_prl}, and the normalized growth rates are given in Fig.~\ref{fig_highmach}. In the subsonic regime, their results indicate that the growth rate (normalized by the eddy-turnover time $T_{ed}$ on the forcing scale) strongly decreases with decreasing Mach number for compressive driving, while it is almost constant at solenoidal driving. At $\mathcal{M}>1$, there is an initial drop due to the appearance of shocks, but increases as $\mathcal{M}^{1/3}$ at larger values. A similar dependence is reported on the saturation level, which is particularly high for solenoidal driving, and decreases in the regime of large, supersonic Mach numbers.

\begin{table}[htdp]
\begin{center}
\begin{tabular}{c|c|c|c|c}
 & $\Gamma_{sol} [T_{ed}^{-1}]$ & $\Gamma_{comp} [T_{ed}^{-1}]$ & $(E_m/E_k)_{sol}$  & $ (E_m/E_k)_{comp}$\\
\hline
$p_0$ & $-18.71$  & $2.251$  & $0.020$  & $0.037$\\
$p_1$ & $0.051$  & $0.119$  & $2.340$  & $1.982$\\
$p_2$  & $-1.059$  & $-0.802$ & $23.33$ & $-0.027$\\
$p_3$ & $2.921$ & $25.53$ & $2.340$ & $3.601$\\
$p_4$ & $1.350$ & $1.686$ & $1$ & $0.395$\\
$p_5$ & $0.313$ & $0.139$ & $0$ & $0.003$ \\
$p_6$ & $1/3$  & $1/3$  & $0$ & $0$
\end{tabular}
\end{center}
\caption{Fit coefficients reported by \citet{FederrathPRL}.}
\label{table_prl}
\end{table}%

We thus summarize the results from numerical simulations as follows:\begin{itemize}
\item The critical magnetic Reynolds number for dynamo action as well as the resulting spectra for the magnetic field depend on the magnetic Prandtl number.
\item Both the growth rates and the saturation levels of the dynamo depend significantly on the turbulent Mach number and the type of forcing that is employed.
\end{itemize}

\section{Non-universality in the non-linear regime}
The exponential growth phase will come to an end when the tension force of the magnetic field, $\vec{B}\cdot\nabla\vec{B}$, becomes comparable to the inertial term of the flow, $\vec{u}\cdot\nabla\vec{u}$. At this point, magnetic field amplification will stop on the scales that fulfill this condition, and continue to proceed on larger scales. As discussed by \citet{Scheko02}, this condition translates to\begin{equation}
\frac{B_{\ell_a}^2}{\ell_a}\sim\frac{u_{\ell_a}^2}{\ell_a},
\end{equation}
where $\ell_a$ denotes the smallest scale where amplification still occurs. In this regime, a linear growth of the magnetic energy has been reported in previous studies, based on the assumption of Kolmogorov turbulence \citep[e.g.][]{Scheko02, Cho09, Beresnyak12}. In the following, we will generalize these investigations by employing a simplified toy model as well as a more sophisticated Fokker-Planck model previously suggested by \citet{Scheko02}. As a result, we will show that different types of power-law growth can be expected depending on the adopted type of turbulence.

We further point out that in the non-linear regime, we expect the magnetic Prandtl number to play a less critical role, as the amplification scale of the magnetic field is now expected to be larger than both the viscous and the resistive scale, such that no strong dependence on Re or Rm can be expected. 

{ We note that the models considered in this section have previously been motivated in the context of the incompressible induction equation, given as
\begin{equation}
\partial_t\vec{B}+\vec{v}\cdot\nabla\vec{B}=\vec{B}\cdot\nabla\vec{v}+\eta\Delta\vec{B}.\label{inductioninc}
\end{equation}
However, they can be naturally extended into the compressible regime with the replacement\begin{equation}
\vec{B}\rightarrow\frac{\vec{B}}{\rho}.
\end{equation}
Inserting this replacement as well as the continuity equation, \begin{equation}
\dot{\rho}=-\nabla\cdot\left(\rho \vec{v}\right),
\end{equation}
it is straightforward to show that one obtains the compressible form of the induction equation,\begin{equation}
\partial_t\vec{B}+\vec{v}\cdot\nabla\vec{B}=\vec{B}\cdot\nabla\vec{v}-\vec{B}\left(\nabla\cdot\vec{v}\right)+\eta\Delta\vec{B},\label{inductioncomp}
\end{equation}
equivalent to Eq.~(\ref{induction}). As long as the mean density $\langle\rho\rangle$ in the box is constant, a significant growth of the quantity $\langle B/\rho\rangle$ nevertheless implies a corresponding growth of the magnetic energy, assuming that the density distribution function will not change significantly over time. In the case of well-developed driven turbulence, one indeed expects a characteristic log-normal density probability distribution function, which naturally complies with these requirements \citep{Vazquez94, Passot98, Federrath08}. Strictly speaking, the following considerations apply to the quantity $\tilde{B}=B/\rho$ and $\tilde{W}=W/\rho^2$, with $W$ the magnetic energy. In the following, the $\tilde{\ }$ is however dropped for simplicity.
}

\subsection{First considerations based on a toy model}\label{toymodel}
In the toy model previously proposed by \citet{Scheko02}, the dominant fraction of the magnetic energy resides on the scale $\ell_a$, the smallest scale where magnetic field amplification still occurs (thus yielding the shortest amplification timescale). On that scale, the magnetic energy is expected to be already close to saturation. The magnetic energy $W(t)$ can thus be related to the amplification scale $\ell_a$ by the approximate relation
\begin{equation}
W(t)\sim\frac{1}{2}\langle\rho\rangle u_{\ell_a(t)}^2.\label{defmagenergy}
\end{equation}
The magnetic energy is evaluated here at the mean density $\langle\rho\rangle$ of the turbulent box, as we are interested only in the magnetic field amplification by shear. Adopting the eddy-turnover rate on the scale $\ell_a$ as the growth rate for the magnetic field, i.e.\begin{equation}
\Gamma(t)\sim\frac{u_{\ell_a(t)}}{\ell_a(t)},\label{Gamma}
\end{equation}
the magnetic energy evolves as\begin{equation}
\frac{d}{dt}W=\Gamma(t)W(t)-2\eta k_{rms}^2W(t)\label{magenergy}
\end{equation}
with \begin{equation}
k_{rms}^2(t)=\frac{1}{W}\int_0^\infty dk k^2M(t,k)
\end{equation}
and\begin{equation}
M(t,k)=\frac{1}{2}\int d\Omega_{\vec{k}}\langle | \vec{B}(t,\vec{k}) |^2\rangle.
\end{equation}
Now, we have $\Gamma(t)W(t)\sim \langle\rho\rangle u_{\ell_a(t)}^3/\ell_a(t)=:\epsilon(t)$. Inserting in Eq.~(\ref{magenergy}) yields\begin{equation}
\frac{d}{dt}W=\chi \epsilon(t)-2\eta k_{rms}^2(t) W(t),
\end{equation}
where $\chi$ is a constant of order unity. For Kolmogorov turbulence, the quantity $\epsilon(t)=\langle\rho\rangle u_{\ell_a(t)}^3/\ell_a(t)$ is a constant \citep{Kolmogorov41}. In this case, and as long as magnetic energy dissipation is negligible, $dW/dt\,=\,\mathrm{const}$, implying a phase of linear growth. In this limit, we  obtain the result of \citet{Beresnyak12}, where a constant fraction of the turbulence dissipation rate is converted into magnetic energy.

In the general case with $u_{\ell_a}\propto \ell_a^\vartheta$, $\epsilon(t)$ is however not constant, but varies as $\ell_a^{3\vartheta-1}$. In the case of Burgers turbulence, we thus obtain $\epsilon\propto \ell_a^{0.5}$.  In this case, the growth of the magnetic energy is no longer linear, as the turbulent energy dissipation rate is not independent of scale!

For comparison, we note that the quantity $\tilde{\epsilon}=\rho e_{\mathrm{SGS}}^{3/2}/\ell$, with $e_{\mathrm{SGS}}$ the specific energy density of subgrid-scale turbulence, is practically independent of $\ell$. It however has a weak dependence on the Mach number, and a strong dependence on the type of forcing \citep{Schmidt11}. As the density fluctuations will however not contribute to the shearing, we will  adopt $\epsilon$ as the quantity of interest here.

To quantify the expected behavior, we need to solve Eq.~(\ref{defmagenergy}) for $\ell_a$. For this purpose, we recall that $u_{\ell_a}$ is related to the turbulence driving scale $L$ and the velocity $V$ on that scale via \begin{equation}
u_{\ell_a}=V\left( \frac{\ell_a}{L}  \right)^\vartheta.
\end{equation}
From (\ref{defmagenergy}), we thus obtain\begin{equation}
\ell_a=L\left( \frac{2W}{\langle\rho\rangle V^2} \right)^{1/(2\vartheta)}.
\end{equation}
We can now evaluate (\ref{Gamma}) and (\ref{magenergy}), yielding\begin{equation}
\frac{d}{dt}W\sim W\left[ L\left( \frac{2W}{\langle\rho\rangle V^2} \right)^{1/(2\vartheta)}  \right]^{\vartheta-1}\propto W^{1+(\vartheta-1)/(2\vartheta)}.
\end{equation}
For Kolmogorov turbulence $(\vartheta=1/3)$, we  confirm that $dW/dt\,=\,\mathrm{const}$, while in the more general case, this quantity will increase with increasing $W$. This can be intuitively understood, as the steep spectra for $\vartheta>1/3$ imply a more modest increase of the eddy-timescale with length scale, suggesting that the amplification rate remains larger when increasing the scale. We re-assess these results with the Fokker-Planck model below and explore the physical implications in more detail.

\subsection{Implications of the Fokker-Planck model for universality}
The starting point for our investigations is the Fokker-Planck model of \citet{Scheko02}. Here, the time-evolution of the magnetic-energy spectrum is given as\begin{equation}
\partial_t M=\frac{\partial}{\partial k}\left[ D(k)\frac{\partial M}{\partial k}-V(k)M  \right]+2\Gamma(t) M-2\eta k^2 M,
\end{equation}
with the diffusion coefficient $D(k)=\Gamma(t) k^2/5$ and the drift velocity in k-space $V(k)=4\Gamma(t) k/5$. We recall that the magnetic-energy spectrum $M$ is related to the magnetic energy $W$ via\begin{equation}
W(t)=\int_0^\infty dk M(t,k).
\end{equation}
To describe the evolution in the nonlinear regime, \citet{Scheko02} postulated the following expressions:\begin{eqnarray}
\Gamma(t)&=&c_1\left[ \int_0^{k_s(t)}dk k^2 E(k)  \right]^{1/2},\label{GammaDef}\\
W(t)&=&c_2\int_{k_s(t)}^\infty dk E(k).\label{satscale}
\end{eqnarray}
The constants $c_1$ and $c_2$ are of order unity, $E(k)$ is the hydrodynamic energy spectrum neglecting the influence of the magnetic field, and the wave vector $k_s(t)$ is defined via Eq.~(\ref{satscale}). It corresponds to the smallest scale where amplification efficiently occurs. As input for the Fokker-Planck model, we require an energy spectrum of the turbulence. As before, we assume that the velocity in the inertial range scales as\begin{equation}
u_\ell\propto \ell^\vartheta.
\end{equation}
The hydrodynamic energy spectrum is then approximately given as\begin{equation}
E(k)=\begin{cases}  C_t \epsilon^{2/3}k^{-2\vartheta-1}  & \mathrm{for\ } k\in[k_f,k_\nu]\\ 0 & \mathrm{elsewhere},      \end{cases}\label{turbspec}
\end{equation}
with $C_t$ a constant which depends on the type of turbulence, $k_f$ and $k_\nu$ the wave vectors describing the injection scale of turbulence and the viscous scale, respectively. The value of $k_\nu$ is set to enforce the condition $\epsilon=2\nu\int_0^\infty  dk k^2 E(k)$. Unlikely in (\ref{TL}) and (\ref{TN}), we do not explicitly model the turbulent spectra in the viscous regime, as these no longer contribute during the non-linear stage. With these input data, Eq.~(\ref{satscale}) can be evaluated as \begin{equation}
W(t)=\frac{c_2 C_t \epsilon^{2/3}}{2\vartheta}\left[ k_s^{-2\vartheta}-k_\nu^{-2\vartheta}  \right].
\end{equation}

We further introduce the quantities \begin{eqnarray}
W_0&=& c_2\int_0^\infty dk E(k)=\frac{c_2 C_t \epsilon^{2/3}}{2\vartheta}\left[ k_f^{-2\vartheta}-k_\nu^{-2\vartheta}  \right],\label{entot}\\
W_\nu&=&\frac{c_2 C_t \epsilon^{2/3}}{2\vartheta}k_\nu^{-2\vartheta}.\label{envis}
\end{eqnarray}
We note that in the above expressions, the integral $\int_0^\infty dk$ corresponds to an integration from $k_f$ to $k_\nu$, as the turbulent energy is non-zero only in this regime (see~\ref{turbspec}).

Using these definitions, the wave vectors $k_\nu$, $k_s$ and $k_f$ can be expressed as\begin{eqnarray}
k_s&=&\left( \frac{2\vartheta}{c_2 C_t \epsilon^{2/3}} \right)^{-1/(2\vartheta)}\left[ W(t)+W_\nu  \right]^{-1/(2\vartheta)},\label{wavevec1}\\
k_f&=&\left( \frac{2\vartheta}{c_2 C_t \epsilon^{2/3}} \right)^{-1/(2\vartheta)}\left[ W_0+W_\nu  \right]^{-1/(2\vartheta)},\label{wavevec2}\\
k_\nu&=&\left( \frac{2\vartheta}{c_2 C_t \epsilon^{2/3}} \right)^{-1/(2\vartheta)} W_\nu^{-1/(2\vartheta)}.\label{wavevec3}
\end{eqnarray}
Integrating Eq.~(\ref{GammaDef}) now yields the following:\begin{equation}
\Gamma(t)=c_1\left[ \left( \frac{C_t\epsilon^{2/3}}{2-2\vartheta} \right) \left( k_s^{2-2\vartheta}(t)-k_f^{2-2\vartheta} \right) \right]^{1/2}.\label{Gammaint}
\end{equation}
Substituting Eqs.~(\ref{wavevec1})-(\ref{wavevec3}) into (\ref{Gammaint}) yields the expression\begin{eqnarray}
\Gamma(t)&=&c_1\left( \frac{C_t\epsilon^{2/3}}{2-2\vartheta} \right)^{1/2} \left( \frac{c_2 C_t \epsilon^{2/3}}{2\vartheta}  \right)^{\frac{1-\vartheta}{2\vartheta}} \\
&\times&\left[ \left(W(t)+W_\nu\right)^{1-\frac{1}{\vartheta}}-\left(W_0+W_\nu\right)^{1-\frac{1}{\vartheta}}  \right]^{1/2}.
\end{eqnarray}
Considering turbulence models between Kolmogorov and Burgers, we have $1/3\leq\vartheta\leq1/2$. We further assume that $W(t)\ll W_0$, implying that the magnetic field is far from saturation on the current amplification scale. In this case, we can neglect the second term in the square brackets. As we focus here on the non-linear regime, we can further neglect $W_\nu$ compared to $W(t)$, and obtain the expression
\begin{equation}
\Gamma(t)=c_1\left( \frac{C_t\epsilon^{2/3}}{2-2\vartheta} \right)^{1/2} \left( \frac{c_2 C_t \epsilon^{2/3}}{2\vartheta}  \right)^{\frac{1-\vartheta}{2\vartheta}}W^{(\vartheta-1)/(2\vartheta)}(t).\label{Gammaapprox}
\end{equation}
As in our toy model, the growth of the magnetic energy  thus scales as\begin{equation}
\frac{d}{dt}W\propto W(t)\Gamma(t)\propto W^{1+(\vartheta-1)/(2\vartheta)}.\label{dWdt}
\end{equation}
For Kolmogorov turbulence, the growth is thus linear, while it grows faster than linear for $\vartheta>1/3$. Integrating Eq.~(\ref{dWdt}), we obtain
\begin{equation}
W(t)=\tilde{C}t^{2\vartheta/(1-\vartheta)},\label{evolutionW}
\end{equation}
with \begin{equation}
\tilde{C}=\left( \frac{C_t\epsilon^{2/3}}{2-2\vartheta} \right)^{1/2}\left( \frac{c_2 C_t\epsilon^{2/3}}{2\vartheta} \right)^{(1-\vartheta)/(2\vartheta)}\left( \frac{5}{2}-\frac{1}{2\vartheta} \right)^{-1}.\label{const}
\end{equation}
From this expression, we already see that the energy  grows linearly in $t$ for Kolmogorov, while it grows as $t^2$ for Burgers turbulence. For a physical interpretation, the normalization in terms of the eddy-turnover time $T_{ed}$ on the forcing scale is still required, which we perform below.

\section{Physical implications}

\begin{figure}[htbp]
\begin{center}
\includegraphics[scale=0.51]{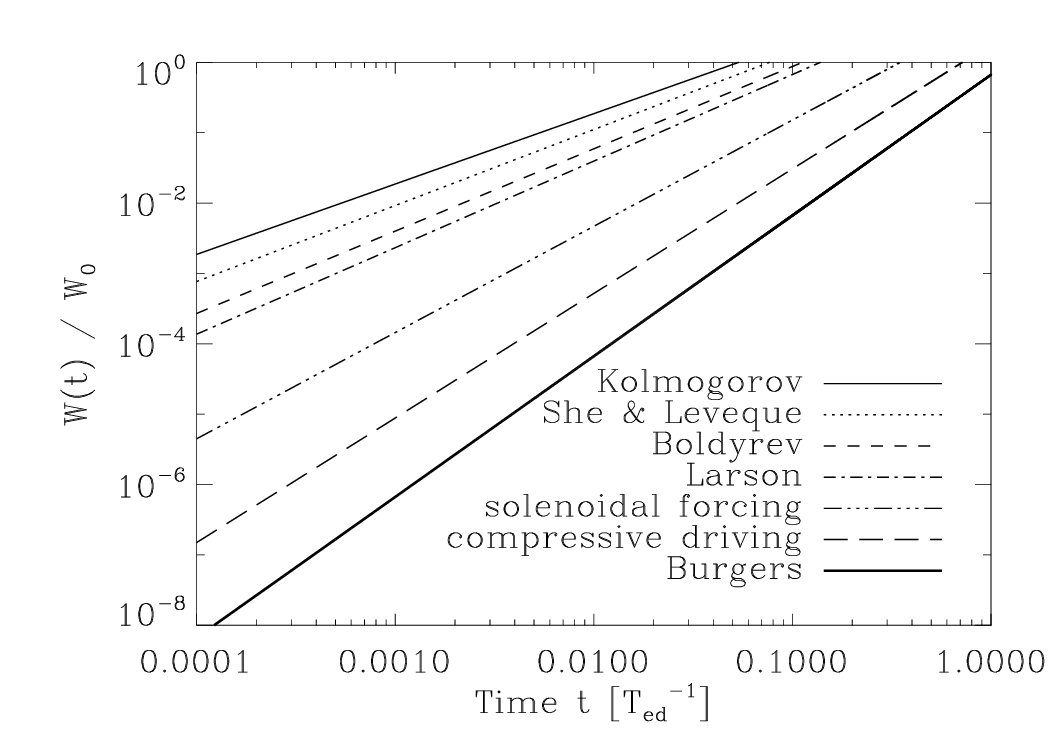}
\caption{The power-law growth of magnetic energy for different types of turbulence in the non-linear regime, following the evolution Eq.~(\ref{evolnorm}) for $\mathrm{Re}=10^4$.}
\label{fig_fieldgrowth}
\end{center}
\end{figure}

\begin{table}[htdp]
\begin{center}
\begin{tabular}{c|c|c|c}
Model and reference & $\vartheta$ & $W\propto$ & $\ell_a\propto$\\
\hline
Kolmogorov \citep{Kolmogorov41}  &  $1/3$ & $t^1$ & $t^{3/2}$\\
Intermittency of Kolmogorov turbulence \citep{She94} & $0.35$ & $t^{1.077}$ & $t^{1.54}$\\
Driven supersonic MHD turbulence \citep{Boldyrev02} & $0.37$   &  $t^{1.17}$ & $t^{1.59}$\\
Observation in molecular clouds  \citep{Larson81} & $0.38$  & $t^{1.23}$ & $t^{1.61}$\\
Solenoidal forcing of turbulence \citep{Federrath10turb} & $0.43$  & $t^{1.51}$ & $t^{1.75}$\\
Compressive forcing of turbulence \citep{Federrath10turb} & $0.47$  & $t^{1.77}$ & $t^{1.89}$\\
Observation in molecular clouds \citep{Ossenkopf02} & $0.47$  & $t^{1.77}$ & $t^{1.89}$\\
Burgers turbulence \citep{Burgers48} & $1/2$  & $t^2$ & $t^2$
\end{tabular}
\end{center}
\caption{The power-law behavior of the small-scale dynamo for different types of turbulence in the non-linear regime.}
\label{tab_pl}
\end{table}%

To explore the physical implications of the above-mentioned results, we now perform a normalization in terms of the eddy-turnover time $T_{ed}$ on the forcing scale $k_f^{-1}$. For this purpose, we note that the expression within the central brackets of Eq.~(\ref{const}) is identical to $W_\nu k_\nu$, and it is straightforward to show that \begin{equation}
W_\nu k_\nu = W_0 k_f \mathrm{Re}^{(1-2\vartheta)/(1+\vartheta)}.
\end{equation}
If we normalize Eq.~(\ref{evolutionW}) in terms of $T_{ed}\sim(k_f \sqrt{W_0})^{-1}$, we thus obtain
\begin{eqnarray}
W(t)&=&C \left( \frac{t}{T_{ed}} \right)^{2\vartheta/(1-\vartheta)},\\
C&=&\left( \frac{C_t\epsilon^{2/3}}{2-2\vartheta} \right)^{1/2}\left( \frac{5}{2}-\frac{1}{2\vartheta} \right)^{-1}\nonumber \\
&\times& \mathrm{Re}^{2\vartheta(1-2\vartheta)/(1+\vartheta)}k_f^{2\vartheta-2\vartheta/(1-\vartheta)}W_0^{2\vartheta-\vartheta/(1-\vartheta)}.\nonumber
\end{eqnarray}
Adopting a system of units with $W_0=1$ and $k_f=1$, it is evident that $E_f\sim1$, $v(k_f)\sim1$ and thus $T_{ed}\sim1$. From Eq.~(\ref{turbspec}), we also expect $\epsilon\sim1$. In these units, our evolution equations simplifies as\begin{eqnarray}
\frac{W(t)}{W_0}&=&C \left( \frac{t}{T_{ed}} \right)^{2\vartheta/(1-\vartheta)},\label{evolnorm}\\
C&=&\left(Ê\frac{1}{2-2\vartheta}Ê\right)^{1/2}\left(\frac{5}{2}-\frac{1}{2\vartheta}  \right)^{-1}\mathrm{Re}^{2\vartheta(1-2\vartheta)/(1-\vartheta)}.\nonumber
\end{eqnarray}
We illustrate the behavior for the different types of turbulence in Fig.~\ref{fig_fieldgrowth} for $\mathrm{Re}=10^4$, and summarize the power-law behavior in Table~\ref{tab_pl}. The solution suggests that turbulence spectra closer to Kolmogorov saturate earlier (in terms of the eddy-turnover time on the forcing scale $k_f$), and initially start at a higher value. The latter is fully consistent with our expectations for the kinematic regime, where the growth rates are higher for Kolmogorov turbulence, and a larger amount of magnetic energy may build up before the non-linear regime is reached (due to the increased amount of turbulent energy that is available on the same scale). We note that in the final stage close to saturation, the evolution may start to deviate from the power-law behavior reported here, providing a transition to the regime where $W(t)=\mathrm{const}$.

From the  relation derived above, we further calculate the characteristic scaling of the current amplification scale $l_s$ as a function of time $t$. Adopting Eq.~(\ref{defmagenergy}), we have $W(t)\sim \langle\rho\rangle u^2_{\ell_a(t)} \propto \ell_a^{2\vartheta}$, thus\begin{equation}
\ell_a(t)\propto W^{1/(2\vartheta)}(t)\propto t^{1/(1-\vartheta)}.
\end{equation}
For Kolmogorov turbulence, the characteristic length scale of the magnetic field thus grows as $t^{3/2}$, while it grows as $t^2$ for Burgers turbulence. The results are summarized for all types of turbulence in Table~(\ref{tab_pl}).

\begin{figure}[htbp]
\begin{center}
\includegraphics[scale=0.5]{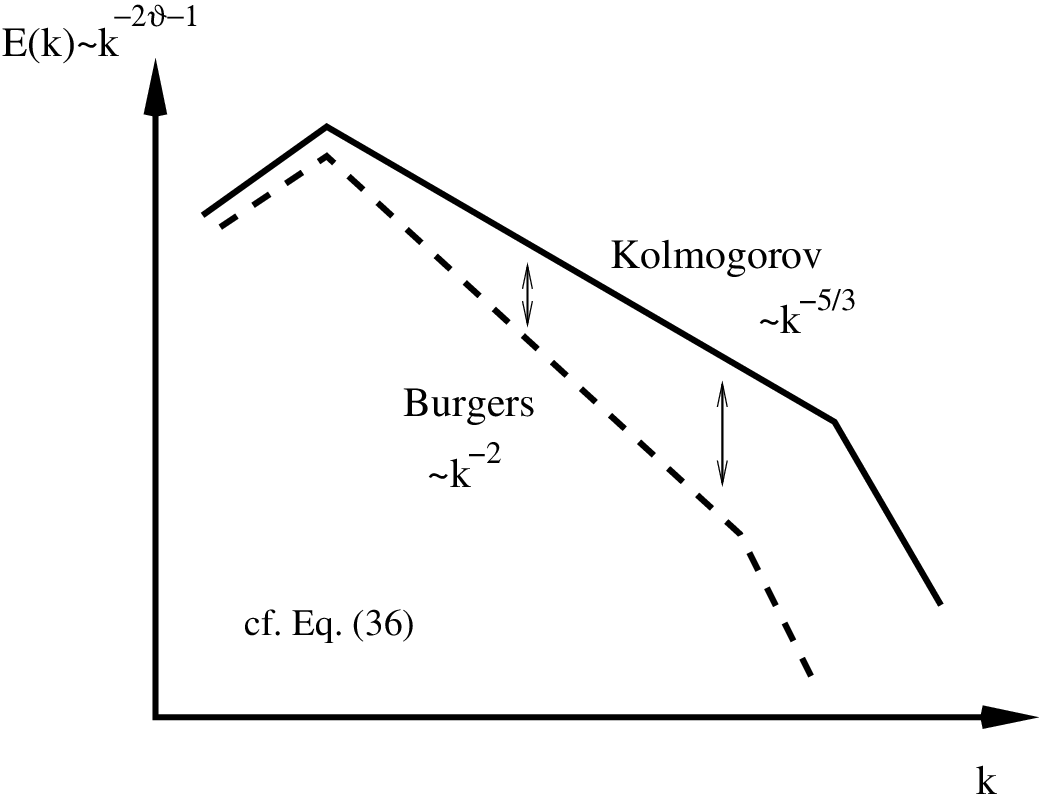}
\caption{A sketch of Kolmogorov vs Burgers turbulence. While the turbulent energy is considerably smaller for Burgers spectra ($\vartheta=1/2$) on small scales, it approaches the values for Kolmogorov turbulence ($\vartheta=1/3$) on larger scales. As a result, the magnetic energy grows faster than linear for Burgers turbulence, as the growth rates gradually approach the Kolmogorov values at later times.}
\label{fig_kolmo_burgers}
\end{center}
\end{figure}

The power-laws derived here  depend on the type of turbulence due to the different eddy-turnover timescales as a function of scale, as we sketch in Fig.~(\ref{fig_kolmo_burgers}). We summarize the main ingredients based on the toy model developed in section~\ref{toymodel}:

 Considering a driving scale $L$ with a turbulence velocity $V$ on that scale, the ratio of the eddy-turnover times on scale $l\ll L$ for Kolmogorov and Burgers turbulence is given as\begin{equation}
\frac{t_K}{t_B}=\frac{(\ell/L)^{1-1/3}}{(\ell/L)^{1-1/2}}=\left( \frac{l}{L} \right)^{1/6}.\label{ratioeddy}
\end{equation}
During the growth of the magnetic energy, the relevant length scale however shifts to larger scales. According to Eq.~(\ref{ratioeddy}), the ratio of the eddy timescales approaches unity for $\ell\rightarrow L$. For Burgers turbulence, the magnetic field amplification is thus initially delayed with respect to Kolmogorov, and catches up later, resulting into the non-linear behavior and the power-law growth described here.

Due to these results, it is clear that the growth rate of the dynamo is not a fixed fraction of the global turbulence dissipation rate, as previously proposed by \citet{Beresnyak12}. Due to the dependence on the turbulent spectrum, such a consideration may only hold locally, i.e. on a given scale, where the growth rate of the field is indeed related to the local eddy timescale. From a more global perspective, however, the turbulence dissipation rate changes as a function of scale for models different from Kolmogorov, such that the previously postulated universal behavior cannot be expected. From Eq.~(\ref{evolnorm}), it is further evident that the evolution depends on the Reynolds number of the gas, and that larger Reynolds numbers imply stronger magnetic fields at earlier times.

\section{Discussion and conclusions}
In this paper, we have explored both the kinematic regime of the small-scale dynamo, where an exponential growth of the magnetic energy is generally observed, and the non-linear regime, where backreactions start occuring on small scales and shift the amplification scale of the magnetic field to larger scales.

In the kinematic regime, analytical studies based on the Kazantsev model suggest a fundamental dependence on the magnetic Prandtl number. In particular, for $\mathrm{Pm}\ll1$, the growth rate of the dynamo is a function of the magnetic Reynolds number Rm, while for $\mathrm{Pm}\gg1$, it depends on the kinematic Reynolds number Re. In addition, the amplification rates significantly depend on the adopted type of turbulence. For $\mathrm{Pm}\gg1$, it scales as Re$^{1/2}$ for Kolmogorov turbulence and as Re$^{1/3}$ for Burgers turbulence. The same scaling relations, with a different normalization, were found for $\mathrm{Pm}\ll1$, with the replacement Re$\rightarrow$Rm.

Numerical simulations confirm the dependence on Pm also in the range $\mathrm{Pm}\sim1$, and find a strong dependence of the growth rate and the saturation level on the turbulent Mach number $\mathcal{M}$ and the type of turbulence forcing. Magnetic field amplification is particularly efficient for solenoidal forcing and low Mach numbers, but  also occurs for high Mach numbers and solenoidal / compressive forcing. If the Mach numbers are very small, compressive forcing is hardly able to trigger magnetic field amplification, as the presence of density gradients are required for the production of solenoidal turbulence in this case.

To investigate the non-linear regime of the dynamo, we employed the Fokker-Planck model of \citet{Scheko02} and explored the effect of different turbulent spectra on the magnetic field amplification rate. We find that the previously known linear growth only occurs for Kolmogorov turbulence, while in the general case with $u_\ell\propto \ell^{\vartheta}$, we expect the magnetic energy to scale as $t^{2\vartheta/(1-\vartheta)}$. The energy growth is thus faster than linear, and may even become quadratic for Burgers turbulence ($\vartheta=1/2$). However, we note that the growth rate is initially smaller for Burgers turbulence, as the turbulent energy available for amplification is initially much smaller on small scales. While magnetic field amplification is shifted to larger scales, the difference in the turbulent energy decreases, implying the reported power-law behavior as a function of time.

We have further shown that also the scaling of the characteristic length scale $\ell_a$ for magnetic field amplification depends on the turbulent slope. Specifically, we find a scaling as $t^{1/(1-\vartheta)}$, corresponding to $t^{3/2}$ for Kolmogorov and $t^{2}$ for Burgers turbulence. The change of length scales proceeds thus in a fashion analogous to the inverse-cascade in case of helicity \citep[e.g.][]{Christensson01, Banerjee04b}. The evolution of this quantity may thus provide another relevant diagnostic for a comparison with numerical simulations.

Due to the above considerations, we point out that the non-linear stage of the small-scale dynamo does not generally correspond to converting a fixed fraction of the turbulence dissipation rate into magnetic energy, as previously suggested by \citet{Beresnyak12}. While their results agree with our model for the case of Kolmogorov turbulence (low Mach numbers),  steeper power laws may occur in the highly compressible regime. Universality in the sense of a uniform behavior under all conditions can thus not be expected. Nevertheless, we note that there are still universal laws governing the behavior of the dynamo, which relate the growth of the magnetic energy to the eddy-turnover time on the current amplification scale. This quantity in general does depend on the Mach number and the type of turbulence involved, such that the breaking of universality is a result of the properties of different environments. { We propose to explore such effects in further detail with numerical simulations to  improve our understanding of such non-universal behavior.}

%We note that our analytic treatment for the high-Mach number regime is of course only approximate, as a fully-selfconsistent treatment including the density fluctuations requires a complete numerical solution. It nevertheless provides a strong indication that universality is not expected from the currently available models, and that high-resolution numerical simulations need to be pursued for a conclusive investigation of the power-law growth during the non-linear stage at high Mach numbers.

\acknowledgements
We thank Robi Banerjee and Ralf Klessen for stimulating discussions on the topic. D.R.G.S., J.S. and S.B. acknowledge funding from the {\em Deutsche Forschungsgemeinschaft} (DFG) in the {\em Schwerpunktprogramm} SPP 1573 ``Physics of the Interstellar Medium" under grant KL 1358/14-1 and SCHL 1964/1-1. D.R.G.S. and W.S.~thank for funding via the SFB 963/1 on ``Astrophysical flow instabilities and turbulence". J.S. acknowledges the support by IMPRS HD, the HGSFP and the SFB~881 ''The Milky Way System''. C.F.~thanks for funding provided by the Australian Research Council under the Discovery Projects scheme (grant DP110102191). { We thank the anonymous referees for valuable suggestions that improved the manuscript.}

\bibliography{astro}
\bibliographystyle{apsrev4-1}

\end{document}